\documentclass{iau}
\usepackage{graphicx}

\title[NOAA 11035] 
{The Persistence of Apparent Non-Magnetostatic \\ Equilibrium in NOAA 11035}

\author[Sarah A. Jaeggli]   
{S. A. Jaeggli}

\affiliation{Department of Physics, Montana State University, \\ P.O. Box 173840, Bozeman, MT 59717-3840, USA \\ email: {\tt jaeggli@solar.physics.montana.edu}}

\pubyear{2015}
\volume{305}  
\jname{Polarimetry: From the Sun to Stars and Stellar Environments}
\editors{K.N. Nagendra, S. Bagnulo, \\ R. Centeno, \& M. Mart\'inez Gonz\'alez, eds.}
\begin{document}

\maketitle

\begin{abstract}
NOAA 11035 was a highly sheared active region that appeared in December 2009 early in the new activity cycle.  The leading polarity sunspot developed a highly unusual feature in its penumbra, an opposite polarity pore with a strong magnetic field in excess of 3500 G along one edge, which persisted for several days during the evolution of the region.  This region was well observed by both space- and ground-based observatories, including Hinode, FIRS, TRACE, and SOHO.  These observations, which span wavelength and atmospheric regimes, provide a complete picture of this unusual feature which may constitute a force-free magnetic field in the photosphere which is produced by the reconnection of magnetic loops low in the solar atmosphere.

\end{abstract}

\firstsection 

\section{Introduction}
In general, sunspot magnetic fields are strongest and vertical in the umbra of a sunspot and weaker and horizontal in the penumbra.  The largest sunspot magnetic fields based on direct measurement of the intensity of Zeeman split line profiles were found in the umbrae of very large, complex sunspots with strengths of up to 6000 G (\cite[Livingston et al. 2006]{livingston06}).  Magnetic fields of up to 7000 G have been inferred from measurements of high-velocity penumbral downflows based on Fourier interpolation and inversion of Stokes profiles (\cite[van Noort et al. (2013)]{vannoort13}), but these occur in tiny patches within the penumbra.  Although it is not a typical $\delta$-spot, the situation in NOAA 11035 bears the greatest similarity to the sunspots observed in \cite[Tanaka (1991)]{tanaka91} and \cite[Zirin \& Wang (1993)]{zirin93}, who found that strong fields were present at the neutral line between opposite polarities in complex sunspots.  However, the observations of the magnetic field and evolution of NOAA 11035 are far more comprehensive than those presented in the work of \cite[Tanaka (1991)]{tanaka91} and \cite[Zirin \& Wang (1993)]{zirin93}.

\section{Observations}
NOAA 11035 was one of the largest sunspot groups produced early in solar activity cycle 24 and was well observed with many space-based instruments, including the Hinode satellite, the Transition Region and Coronal Explorer (TRACE), and the Solar and Heliospheric Observatory (SOHO).  During the same time period I obtained ground-based observations with the Dunn Solar Telescope Facility Infrared Spectropolarimeter (FIRS) and the Interferometric Bidimensional Spectrometer (IBIS).

\begin{figure}[t]
	\begin{center}
		\includegraphics[width=5.4in]{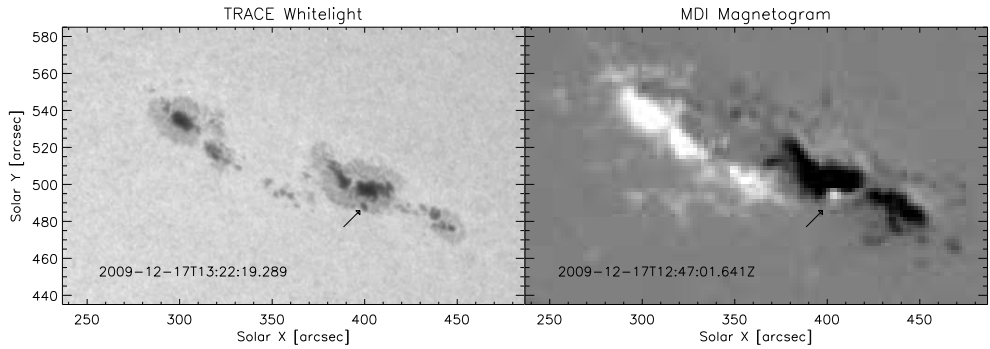}
		\caption{TRACE whitelight (left) and MDI LOS magnetogram (right) taken nearest in time to the Hinode SOT/SP scan near 13:30 UT on 2009-12-17.  The strong magnetic feature is indicated by the red arrow and clearly corresponds to a region with opposite polarity.}
		\label{fig1}
	\end{center}
\end{figure}

The active region emerged near the central meridian on Dec 14, 2009, early in solar cycle 24.  In accordance with Joy's law (\cite[Hale et al. 1919]{hale19}) this region appeared at high latitude (510'' north and 300'' east of disk center), had a large inclination to the solar equator, and the magnetic field was highly sheared.  The region evolved rapidly over the next few days and decayed to a simple sunspot pair as it reached the limb on Dec 20.  Figure \ref{fig1} shows how the region looked on Dec 17 around 13:30 UT.  A movie showing the full evolution of the region from Dec 15-20 can be found in the electronic version supporting material.

I limit the scope of this work to a subfield of the Fe I 6302 \AA\ spectropolarimetric raster obtained with the Hinode Solar Optical Telescope Spectropolarimeter (SOT/SP, \cite[Lites et al. 2013]{lites13}) from 13:30:05 to 13:54:20 on Dec 17, 2009.  The fast map SOT/SP observations have a spatial sampling of 0.30''$\times$0.32'' and spectral resolution of 25 m\AA.  Observations from the TRACE white light imager (\cite[Handy et al. 1999]{handy99}) with 1'' spatial resolution and line of sight magnetic flux measurements from SOHO's Michelson-Doppler Imager (\cite[Scherrer et al. 1995]{scherrer95}) with 4'' spatial resolution were taken with a 60 minute cadence throughout the lifetime of the region and provide excellent context.  A He I 10830 \AA\ spectroscopic raster was obtained with FIRS (\cite[Jaeggli et al. 2010]{jaeggli10}, \cite[Jaeggli 2011]{jaeggli11}) from 16:43 to 17:06 on Dec 17, 2009 and provides useful information on the chromospheric structure above the region.  FIRS has 70 m\AA\ spectral resolution in the infrared and can achieve 0.36'' spatial resolution, although the seeing conditions were deteriorating throughout the scan.

Reduced level 1 data from Hinode SOT/SP was obtained from the Community Spectro-polarimetric Analysis Center archive.  Level 1 TRACE and MDI data were taken from the archive at Montana State University.  FIRS data were corrected for dark current, detector non-linearity, and flat field variations with the same techniques described in \cite[Jaeggli, Lin, \& Uitenbroek (2012)]{jaeggli12}.  All images and raster maps from all instruments have been oriented such that solar north is up and east is left.  There is a systematic $180^\circ$ difference in the line of sight magnetic field inclination between MDI and SOT/SP.

\section{Analysis of Spectropolarimetry}
\begin{figure}[t]
	\begin{center}
		\includegraphics[width=3.5in]{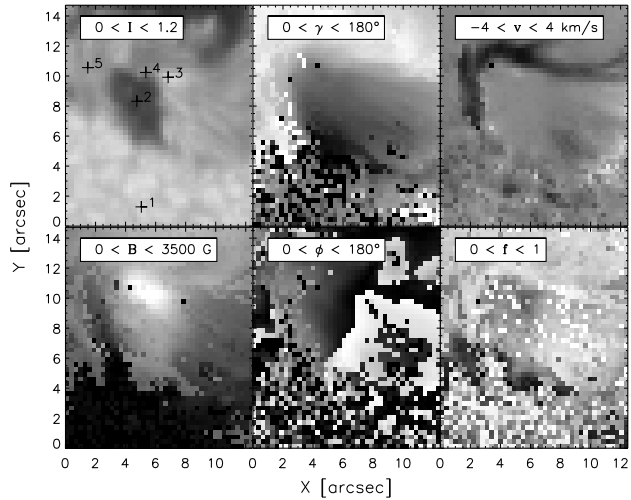}
		\caption{Maps of results from the 2CMO inversion code using one magnetic, and one non-magnetic component.}
		\label{fig2}
	\end{center}
\end{figure}

The Two Component Magneto-Optical (2CMO) Milne-Eddington inversion code was used to infer the vector magnetic field from the SOT/SP full-Stokes spectra.  This code is implemented in IDL and includes the ability to interactively invert profiles.  The previous version, with one magnetic component and a stray light profile, produced results nearly identical to the MERLIN inversion code.  This code is based heavily on the formalism of \cite[Jefferies, Lites, \& Skumanich (1989)]{jefferies89}, and the interested reader should refer to this paper for more details.  I have updated this code from the version described in \cite[Jaeggli, Lin, \& Uitenbroek (2012)]{jaeggli12} with an improved initialization for the magnetic field strength following the center of gravity technique of \cite[Rees \& Semel (1979)]{rees79}, and for the magnetic field azimuth and inclination angles following \cite[Auer, Heasley, \& House (1977)]{auer77}, and the ability to handle multiple magnetic, non-magnetic, and stray light components.  The results of a 2CMO inversion of the SOT/SP sub-field with one magnetic component and one quiet-Sun component are shown in Figure \ref{fig2}.

\begin{figure}[t!]
	\begin{center}
		\includegraphics[width=5in]{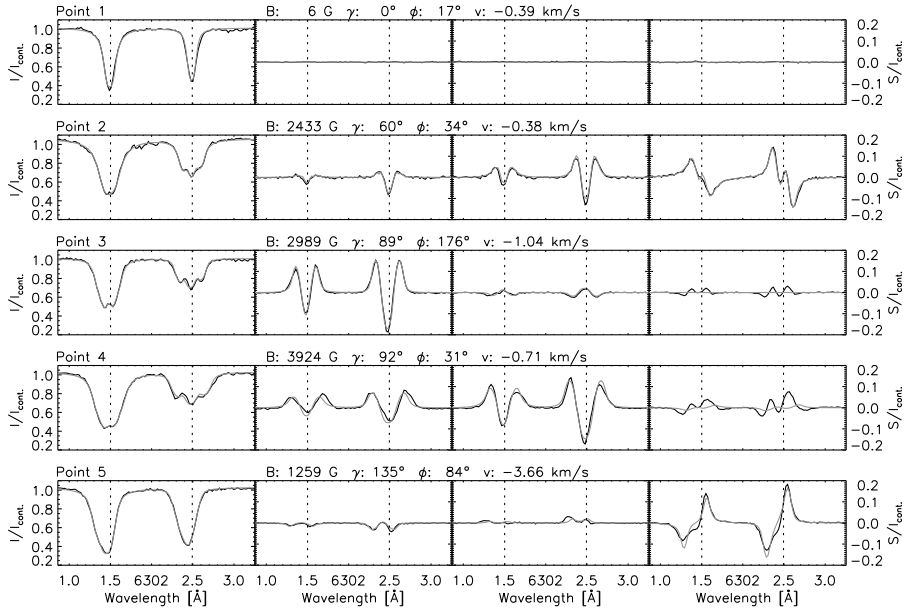} 
		\caption{Profiles from the points indicated in Figure \ref{fig2} fitted using the 2CMO code with one magnetic, and one non-magnetic component.  The black lines are the observed profiles and the grey lines are the fits.}
		\label{fig3}
	\end{center}
\end{figure}

In the SOT/SP raster, regions around the upper boundary of the pore show complex spectral profiles in Stokes Q, U, and V, indicating that there are multiple, velocity shifted components which are unresolved.  Figure \ref{fig3} shows the observed profiles in black from a representative selection of positions indicated in the upper left panel of Figure \ref{fig2}.  The third and fourth profiles, taken from the edge and middle of the region which shows the strongest magnetic field, show clear complexity in their Stokes V profiles which are not properly fit by the inversion technique.

\begin{figure}[h!]
	\begin{center}
		\includegraphics[width=5in]{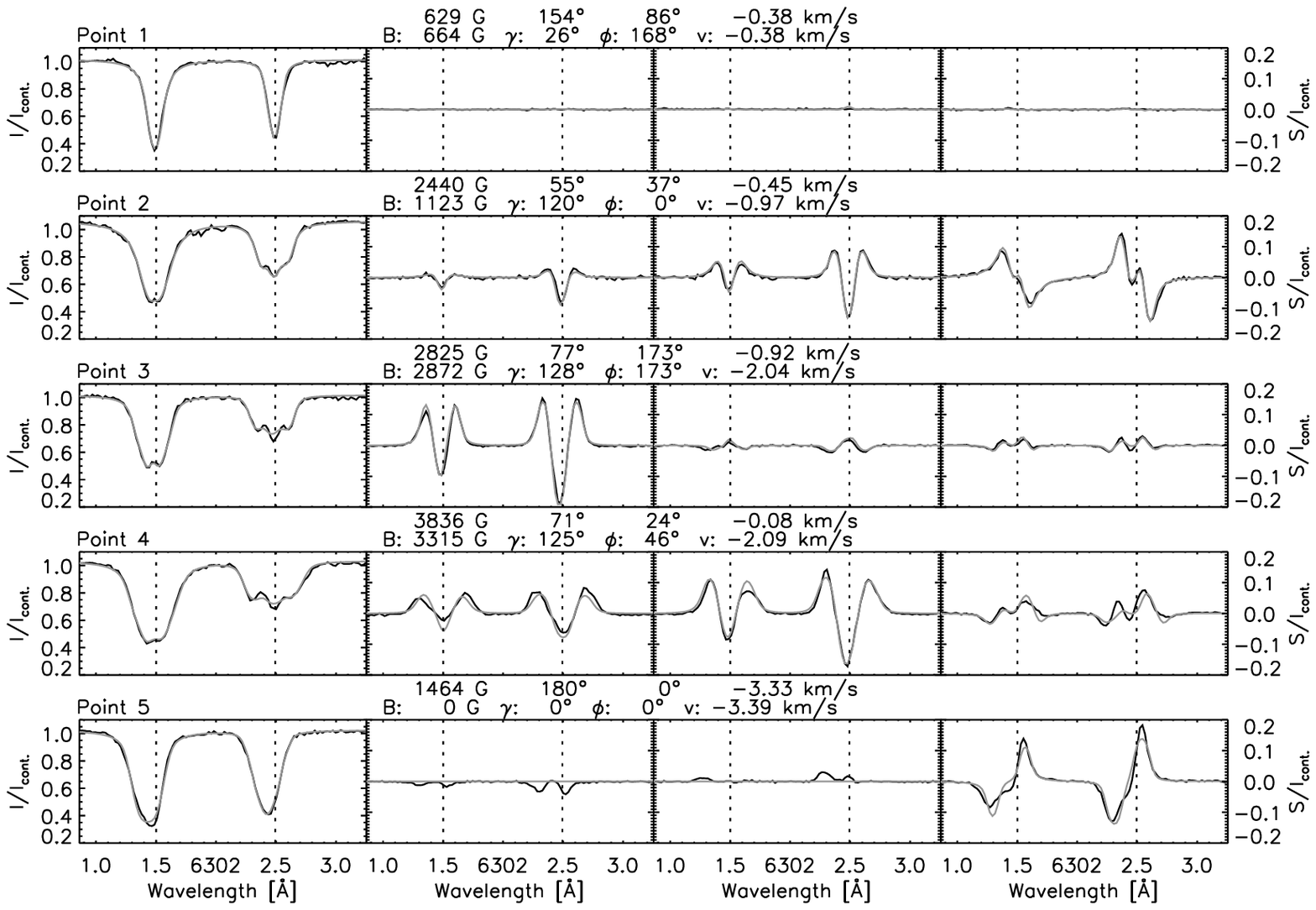} 
		\caption{Profiles from the points indicated in Figure \ref{fig2} fitted using the 2CMO code with two magnetic components. The black lines are the observed profiles and the grey lines are the fits.}
		\label{fig4}
	\end{center}
\end{figure}

In order to fit the complex Stokes V profiles I have eliminated the non-magnetic component and allowed for a second magnetic component.  The two magnetic components are constrained to have the same field strength, but the field orientation and velocity of line center are allowed to be independent.  This produces the fitted profiles shown in Figure \ref{fig4}.  The fits to the Stokes V components for the last three profiles are superior to those using a single magnetic component.  However, the magnetic field strength required to reproduce these profiles is only $\sim200$ G smaller than for a single profile.  This seems to indicate that a strong magnetic field throughout the line of sight is required to produce these profiles.

\section{MHSE Calculation}
Strong, horizontal fields are not generally permitted because they are buoyant.  Magnetic pressure allows the gas density within a tube of magnetic flux to be lower than that of the surrounding atmosphere, causing it to rise (\cite[Parker 1955]{parker55}).  If the magnetic field is deeply embedded in the atmosphere it could be constrained by gas pressure alone, but such depths only become viewable within cool regions like sunspots or bright points in granular lanes (i.e. where the magnetic field is vertical).  A flux tube may also satisfy the conditions for magnetohydrostatic equilibrium if the tube is steeply curved and the magnetic tension force is large.  Consider a simple rectangular ribbon of magnetic flux which is lying horizontally in the photosphere, and assume a simple form for the magnetic field vector which varies along the long dimension of the ribbon ($\hat x$) but does not vary along the vertical thickness ($\hat z$) or horizontal width ($\hat y$) of the ribbon.  If we choose a simple functional form for the magnetic field vector with no radial dependence, a Gaussian-like $B_z(z)$ with a peak at zero, and a sigmoid-like $B_r(z)$ with $B_r(0)=0$ increasing to positive and negative values on each side, the equilibrium state is easy to calculate.  Starting from the general equation for magnetohydrostatic equilibrium:
\begin{equation}\label{eqn1}
\nabla P = {1 \over 8\pi} (\nabla (B^2) + 2 (\vec B \cdot \nabla) \vec B)
\end{equation}
we will assume that gas pressure ($P$) inside and outside of the ribbon is balanced, i.e. the situation is force free, because material can enter the ribbon from the ends.  This leaves the magnetic pressure term and the magnetic tension or curvature force to balance each other.  The symmetry of the assumed magnetic field makes the integral of Equation \ref{eqn1} simple to calculate over the domain of the ribbon and we find that the horizontal field component at the middle of the ribbon ($B_x$) must be equal to the change in the vertical magnetic field over the ribbon ($\Delta B_z$).

\begin{figure}[t]
	\begin{center}
		\includegraphics[height=2.in]{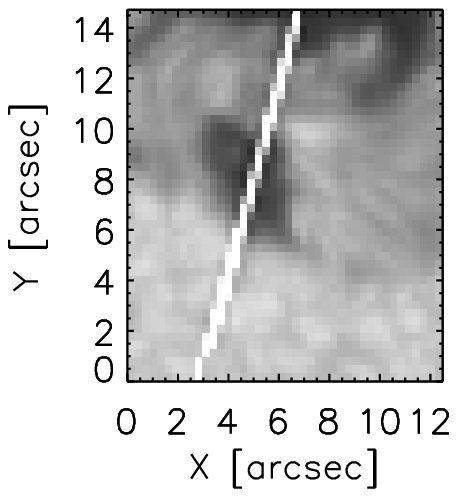}\includegraphics[height=2.in]{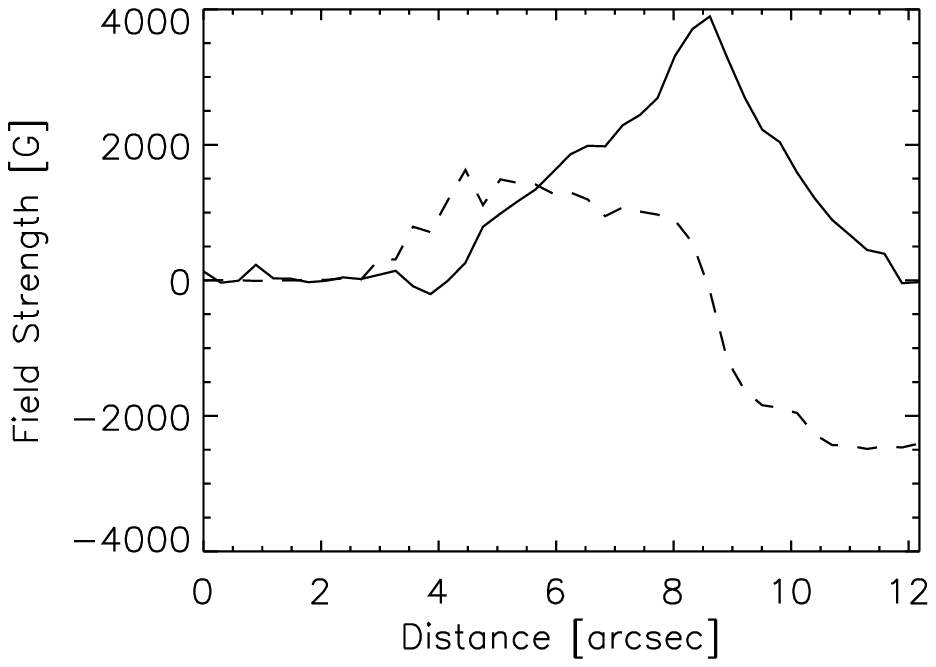} 
		\caption{The profile of the horizontal (solid line) and vertical (dashed line) magnetic field vector along the slice indicated on the left.}
		\label{fig5}
	\end{center}
\end{figure}

This result applies particularly well to the case of the magnetic field in NOAA 11035.  Figure \ref{fig5} shows the vertical and horizontal components of the magnetic field in a slice through the pore from south east to north west for results from the inversion with one magnetic and one non-magnetic component.  The horizontal magnetic field peaks at $\sim$4000 G and the vertical magnetic field changes from +2000 G to -2000 G along the slice.  The change in the vertical component of the magnetic field is sufficiently rapid to constrain the very strong horizontal magnetic field without considering a contribution from the mechanical pressure of the gas.

\section{Chromopsheric Structure}
\begin{figure}[t!]
	\begin{center}
		\includegraphics[width=3.5in]{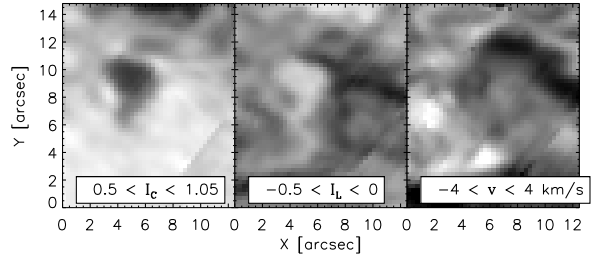}
		\caption{Map of the He I 10830 \AA\ continuum, line amplitude, and velocity from a FIRS observation spanning 16:43 - 17:06 on Dec 17, 2009.}
		\label{fig6}
	\end{center}
\end{figure}

A simple 3-Gaussian fit was applied to the Stokes I component of the He I 10830 \AA\ triplet, using a fixed separation in wavelength and a fixed ratio between the line strengths.  The intensity of the red line of 10830 \AA\ and velocity, along with the intensity of the nearby infrared continuum, are shown in Figure \ref{fig6} for a region around the pore, taken starting at 16:43.  There are some evolutionary differences in the region because this observation is several hours after the SOT/SP scan shown above.  A dark structure is present extending across the upper edge of the pore, and upflowing material is also seen at chromospheric heights, although the position of the upflow is offset conspicuously to the west relative to the Fe I 6302 \AA\ line.

\section{Conclusions}
We have discovered a feature in a sunspot penumbra with a very high magnetic field strength.  We have determined that the asymmetry in the Stokes profiles cannot be reproduced by two weak field components.  Large blue shifts are seen in the photosphere and chromosphere above the neutral line between the pore and the sunspot umbra.  This would seem to indicate that the pore magnetic field is continuously reconnecting with the surrounding active region magnetic field, lofting dense photospheric material into the chromosphere.  If this is the case, the strong magnetic component results from the contraction of the lower loops resulting from reconnection where they pile up against the photosphere.

\begin{discussion}

	\discuss{Solanki}{It's nice to have independent verification of these features}

	\discuss{Jaeggli}{Thanks, I'm interested in seeing other examples of this.}

	\discuss{L\'opez Ariste}{Do you have polarimetric data for the 10830 line?  Have you analyzed this?}

	\discuss{Jaeggli}{Yes, but the quality of the polarimetry data is not very good.}

\end{discussion}

\end{document}